
\magnification=\magstep1
\def \n{\noindent}
\def \v{\vskip 0.5 truecm}
\def \vs{\vskip 1 truecm}

\baselineskip = 18 truept
\def \n {\noindent}
\overfullrule=0pt

\vs
\centerline{\bf ON INDUCED GRAVITY IN 2-d TOPOLOGICAL THEORIES}
\vs
\centerline{\it D.AMATI}
\v
\centerline{S.I.S.S.A.}
\centerline{Scuola Internazionale Superiore di Studi
Avanzati, Trieste}
\centerline{and INFN, Sezione di Trieste}
\vs
\centerline{\it S.ELITZUR and E.RABINOVICI\footnote{*}{Presently at the
Institute for Advanced Studies - Princeton}}
\v
\centerline{S.I.S.S.A.}
\centerline{Scuola Internazionale Superiore di Studi
Avanzati, Trieste}
\centerline{and Racah Institute of Physics, Hebrew University, Jerusalem}

\vskip 3truecm
\centerline{\bf ABSTRACT}

\vs
We study 2-d $\phi F$ gauge theories with the objective to understand,
also at the quantum level, the emergence of induced gravity. The wave
functionals - representing the eigenstates of a vanishing flat potential
- are obtained in the $\phi$ representation. The composition of the
space they describe is then analyzed: the state corresponding to the
singlet representation of the gauge group describes a topological
universe. For other representations a metric which is invariant under
the residual gauge group is induced, apart from possible topological
obstructions. Being inherited from the group metric it is rather
rigid.
\vskip 1.5 truecm
\centerline{Ref. S.I.S.S.A. 160/93/EP}
\centerline{Ref. RI/154(1993)}
\centerline{Ref. IAS-SNS-HEP-93/70}
\vfill
\eject

\n
{\bf A. Introduction.}
\v
It has often been suggested that the difficulties of gravity may be resolved by
the ``well
doings'' of a larger symmetry effective at small distances.  It has been
advocated [1], [2] that
a non-metric - or topological - realization of general coordinate
invariance (GCI) could be a
possibility in this sense.  In these approaches a metric, and with it Einstein
 gravity, would
appear as a large distance effect of a semi-classical nature.  This is
 essentially implemented
by a non-symmetric classical configuration that would spontaneously break the
   higher
symmetry (including a gauge group $G$).  On such configurations a general
coordinate
transformation is implemented by a gauge one;  in this manner the metric on the
group
induces a space-time metric.

This semi-classical approach is in general not sufficient  to investigate the
full
quantum structure
which is essential to understand how these induced metric theories may overcome
the diseases
of usual gravity.  Much effort has been devoted to investigating 2 and 3-d
models - mainly
$CS$ for 3-d [3] and $\phi F$ [4] for 2-d,  where the theory may be solved
 exactly
at the quantum level.  In this paper we reconsider the topological 2-d $\phi F$
 theory,
discussing  in some detail its gauge structure and determining the full quantum
 wave
functional. All wave functionals have the same zero energy as should be in any
 topological
theory;  all these models thus have in common a flat potential structure.

In section B we study gauge theory aspects of the $\phi F$ models.  We solve
 them quantum
mechanically in the $A_0 = 0$ gauge on a space circle (as well as an interval
 for the open
string case).  We discuss compact (simply connected and non simply connected)
as
    well as
non compact groups that are related to gravitational models and in which $\phi$
 was given a
target space coordinate role.

The wave functionals have support only for field configurations for which all
 group invariants
are space independent.  This allows a basis for the wave functionals whose
 members are in
one to one correspondence with the highest weight representations of the group.
 In this basis
each wave functional acquires a phase that has a geometrical meaning.  The same
 phase may
also serve as an action of a quantum mechanical system.  These results are
 connected with
those of a semiclassical limit of the 3-d C.S. theory [5].

The wave functional corresponding to the singlet of the group - and thus
 corresponding to a
vanishing field $\phi$ - plays a special role among the possible vacua allowed
 by the flat
potential.  It describes a $G$ invariant state.  For all other wave functionals
 the gauge group
$G$ is broken down to a residual symmetry group $H$ (generically composed of
 one-dimensional factors).

We also solve for the wave functionals in the presence of sources; the scalar
 field acts as the
electric field does in the strong coupling phase of confining gauge theories.
 With the
difference, however, that the flux string is tensionless.

In section $C$ we turn to the gravitational interpretation of some of these
 results.  The singlet
configuration cannot describe a metric and thus represents a topological phase.
 For the other
wave functionals - corresponding to non-singlet representations of the group
$G$
 - a space
time metric which is a scalar under the residual group $H$ pops out both in
 target space and
in  world sheet.  This is possible as long as the configuration contains fields
 $\phi$ with non-
vanishing spatial derivatives.  A constant field $\phi$ (defining target space
 as a single point)
gives back a non-metric description as expected from the fact that it does not
 allow a
differentiation among different points in space.

We confirm the results obtained in terms of wave functionals by evaluating an
 order
parameter which was suggested [6] to distinguish different phases of gravity.

The induced metric, where it exists, has a rather rigid structure: the constant
 curvature
inherited from the group.  This may be a sign of the limited physical content
of
 these 2-d
models.  In particular,  for the central extension of the $SO(2, 1) \times
U(1)$
  groups [7],
[8],  the H invariant induced metric has zero curvature unlike that for a black
 hole.

\vs
\noindent
{\bf B. General two dimensional $\phi F$ theory-gauge theory aspects.}

\v
\noindent
{\bf B1} {\it The classical $\phi F$ model.}
\v
The topological two dimensional $\phi F$ gauge theory of a Lie group $G$
consists of a gauge connection $A$ in the adjoint representation as well
as $dim (G)$ real scalar fields transforming under the adjoint
representation of $G$.

The topological action is given by:
$$
S = \int_\Sigma Tr(\phi \cdot F)d^2 x\leqno{(B1-1)}
$$
where $F = dA + A^2$ is the field strength corresponding to the gauge
connection $A$ and $tr(XY) = \eta_{ab}X^a Y^b$ is a real invariant bi-linear
form on the Lie algebra of $G$. $\Sigma$ is the two dimensional world
sheet.

The equation of motion are:
$$
F^a_{\mu \nu} = 0\leqno{(B1-2)}
$$
$$
D_\mu \phi^a = 0\leqno{(B1-3)}
$$
where $D_\mu \phi^a$ is the covariant derivative of $\phi$
$$
(D_\mu \phi)^a = \partial_\mu \phi^a + f^a_{bc} A_\mu^b
\phi^c\leqno{(B1-4)}
$$
$f^a_{bc}$ being the structure constants of $G$.

The topological theory can be studied by considering various properties
of the moduli space of the equations of motion [4]. In particular
the partition function measures some appropriately chosen volume of the
moduli space (which depends on the group $G$ and the genus of $\Sigma$).
\v
\noindent
{\bf B2} {\it Quantization of the $\phi \cdot F$ model.}
\v
In this work we perform a canonical quantization in the $A^a_0 = 0$
gauge. From now on $A$ will denote $A_1$.

In order to be able to apply a Hamiltonian formulation we chose
$\Sigma$ to be $R^1 \times S^1$.

{}From eq. (B1-1) it follows that the gauge fields $A$ and the scalar fields
$\phi$ are canonical conjugate to each other, the equal time commutation
relations
they satisfy are:
$$
[\phi^a(x), A^b(y)] = - i\hbar \eta^{ab}\delta(x - y) \leqno{(B2-1)}
$$
we can thus choose a complete set among the $(\phi^a, A^a)$ variables.

Some
of the results of quantization can be anticipated by recalling that the
two dimensional $\phi \cdot F$ theory (eq. (B1-1)) is a dimensionally
reduced 3-dimensional Chern-Simon theory (CS) of the same gauge group
$G$ defined by the action
$$
S = {k \over {4\pi}} \int_M Tr(A \wedge dA - {2 \over 3} A
\wedge A \wedge A)\leqno{(B2-2)}
$$
on a three-manifold $M = \Sigma \times S^1$ ($\Sigma$ is the 2-d
manifold of eq. (B1-1)). The theories described by equations (B1-1) and
(B2-2) are both topological, the reason a scale dependent procedure such
as dimensional reduction retained the topological nature of the theory
is that dimensional reduction results actually by considering the large
$k$ limit of the CS theory in eq. (B2-2), [5].

In that limit, only configurations of the type (B1-2) contribute as long
as $G$ has no center. For a general $G$
$$
Z_{cs}(G, k) \rightarrow  Z_{\phi \cdot
F}(G)exp(a \chi(\Sigma))\# Z(G) \qquad \rm{for} \, k \to \infty\leqno{(B2-3)}
$$
where $a$ does not depend on $\Sigma$, $\chi(\Sigma)$ is the Euler
number of $\Sigma$ and $\#Z(G)$ the number of elements in the center of
$G$. Now, it is known [9] that the Hilbert space of the CS model (B2-2)
consists (on $\Sigma = T^2$) of the conformal blocks of the affine Lie
algebra of $G$. In particular in the large $k$ limit, to each
representation of $G$ corresponds one state in the Hilbert space.

Thus the Hilbert space of the $\phi \cdot F$ theory (B1.1) should also be
in one to one correspondence to the representations of $G$.
We will verify that by an explicit calculation of the Hilbert space. In
addition, in the CS case the phase of the wave functional had an
interesting structure, it was identical to the action of the chiral
$WZNW$
model [9], [10]. In the $\phi \cdot F$ case the phase will also
have a geometrical interpretation.

The $\phi F$ theory being topological requires all wave functionals to
have zero energy. The wave functionals themselves are chosen so that
they are invariant under the residual gauge transformations in the
$A^a_0 = 0$ gauge. This imposes constraints on the states.

It is possible [4] to represent the wave functionals in an $A^a$ basis.
A complete basis for the allowed wave functionals are the characters $Tr
_r(w)$, where $r$ is an irreducible representation of $G$, $w$ is the
Wilson loop constructed from the connection $A$ by:
$$
w = P exp \oint dx' A(x')\leqno{(B2-4)}
$$
where $P$ denotes a path ordered exponential taken around the
circumference of space $S^1$.

For our goal to connect the theory to gravity and to the emergence of a
target space, we find it instructive to construct the wave functional in
the $\phi^a$ basis.

While in CS theories the two possible conjugate coordinates $A^a_1$ and
$A^a_2$ had the same properties under gauge transformations, $A^a$ and
$\phi^a$ in (B2-1) have different transformation properties under the
gauge group $G$. If we denote $\Psi(\phi)$ the wave functional in the
$\phi^a(x)$ basis the operators $A_a(x) = \eta_{ab}A^b(x)$ act upon
$\Psi(\phi(x))$ as $i {\delta \over{\delta \phi^a(x)}}$.

The generators of the residual gauge symmetries are given by the spatial
component of equation (B1-4)
$$
D_1 \phi^a = \partial_1 \phi^a + f^a_{bc} A^b \phi^c .\leqno{(B2-5)}
$$
They are obtained by varying eq. (B1-1) with respect to $A^a_0$. We will
suppress the spatial index in the derivative term.

When expressed in terms of the basis $\phi^a(x)$, the fact that physical
states should obey Gauss's law is represented by:
$$
G^a(x)\Psi(\phi(x)) = [\partial \phi^a(x) + if_b^{ca} \phi^b {\delta
\over {\delta \phi^c}}] \Psi(\phi(x)) = 0 .\leqno{(B2-6)}
$$
Before actually solving eq. (B2-6) note that it constrains the
configurations $\phi(x)$ for which $\Psi(\phi(x))$ does not vanish.

We shall indeed show that $\Psi(\phi(x))$ does not vanish only for
configurations $\phi(x)$ which are of the form
$$
\phi(x) = h^{-1}(x)\phi_D h(x)\leqno{(B2-7)}
$$
where $\phi_D$ are constant in space and belong to the Cartan subalgebra
$C$ of $G$ or rather, to a specific Weyl chamber in $C$. The matrix
$h(x)$ stands for a general element of $G$. In particular, every Casimir
invariant polynomial of the Lie algebra, when calculated in terms of
$\phi(x)$, which are in the support of $\Psi(\phi(x))$, is $x$
independent. This occurs since such $\phi(x)$, are contained in a single
orbit of the adjoint action of $G$. To see that the support of
$\Psi(\phi(x))$ is given by eq. (B2-7), consider for a given $\phi(x)$
another Lie algebra element, $\tau(x)$, which is required to commute
with $\phi(x)$.
$$
[\tau(x), \phi(x)] = 0.\leqno{(B2-8)}
$$
Multiplying eq. (B2-6) by $\tau(x)$ and taking the trace, one obtains:
$$
Tr(\tau(x) \partial \phi(x)) = 0.\leqno{(B2-9)}
$$
For any $\phi(x)$ there exists a group element $h(x)$ such that
$$
\phi_D(x) = h(x) \phi(x)h^{-1}(x)\leqno{(B2-10)}
$$
belongs to a suitable chosen Cartan subalgebra $C$ (this point will be
expanded upon for non compact groups $G$). $h(x)$ is not unique, but
this will be of interest only at a later stage.

Let $\tau_i$, $1 < i < r$, be a basis for $C$, $r$ denotes the rank of
$G$. Since $\tau_i \varepsilon C$ it follows that
$$
[\tau_i, \phi_D(x)] = 0\leqno{(B2-11)}
$$
conjugating by $h(x)$ one has by eq. (B2-9)
$$
Tr(h^{-1}(x)\tau_ih(x)\partial \phi(x)) = Tr(\tau_i h(x)\partial
\phi(x)h^{-1}(x)) = 0\leqno{(B2-12)}
$$
using eq. (B2-10) for $\phi(x)$ it follows from eq. (B2-12) that
$$
0 = Tr (\tau_ih(x)\partial \phi(x)h^{-1}(x)) =
Tr(\tau_ih\partial(h^{-1}\phi_Dh)h^{-1}) =
$$
$$
Tr(\tau_i(\partial \phi_D +
h\partial h^{-1}\phi_D + \phi_D(\partial h h^{-1}))) = Tr(\tau_i\partial
\phi_D).\leqno{(B2-13)}
$$
The element $\partial \phi_D$ belongs to $C$ as well. It can thus be
expanded in the basis $\tau_i$, by eq. (B2-13) all its components vanish,
proving that $\phi_D$ is $x$ independent. One can solve explicitly the
remaining dependence of $\Psi(\phi(x))$ on $\phi(x)$. For example for
the case $G = SU(2)$, one can show that the wave functional
$$
\Psi(\rho, \sigma, \alpha) = exp(\oint {{\rho \partial \sigma - \sigma
\partial \rho}\over{M^2 - \alpha^2}} \alpha dx) \quad
F(M^2)\leqno{(B2-14)}
$$
satisfies eq. (B2-6) where
$$
\eqalign{&\phi = \phi_1\tau_1 + \phi_2\tau_2 + \phi_3\tau_3\cr
&\rho = {{\phi_1 + i\phi_2}\over \sqrt 2}\cr
&\sigma = {{\phi_1 - i\phi_2}\over \sqrt 2}\cr
&\alpha = \phi_3\cr
&M^2 = \phi^2_1 + \phi^2_2 + \phi^2_3 = 2\rho \sigma + \alpha^2}\leqno{(B2-15)}
$$
$\tau_i$ being the Pauli matrices (in this case ${1 \over 2}Tr\phi^2_D = M^2$
an
d thus
$M$ is constant). $F$ is an arbitrary function of $M^2$ (arbitrary up to
some normalization requirement on the wave function).

In order to proceed for a general $G$ it will be useful to be a little
less explicit on the dependence of $\Psi$ on the angular field variables.
The solution will be obtained along the lines of the determination of
the wave functional in CS theory [10] respecting the differences
between the cases.

To solve eq. (B2-6) on admissible configurations (of the type eq. (B2-7))
recall that Gauss' laws are the infinitesimal generators of time
independent gauge transformations. Indeed, under a gauge transformation
$g(x)$, the operators $\phi(x)$  and $A(x)$ transform as:
$$
\phi(x) \to g(x)\phi(x)g^{-1}(x)\leqno{(B2-16a)}
$$
$$
A(x) = i{\delta \over{\delta \phi(x)}} \to g(x) i {\delta \over {\delta
\phi(x)}} g^{-1} + g(x)\partial g^{-1}(x).\leqno{(B2-16b)}
$$
To implement these transformations, the unitary operator, $U_g$, acting
on $\Psi$, corresponding to $g(x)$ should be defined by:
$$
(U_g(x)\Psi)(\phi(x)) = exp(- iTr\int
gdg^{-1}\phi)\Psi(g^{-1}(x)\phi(x)g(x)).\leqno{(B2-17)}
$$
In eq. (B2-17) the transformation of the argument of $\Psi$ gives rise to
the homogeneous part of eq. (B2-16), while the phase factor in front is
needed to produce the inhomogeneous part in eq. (B2-16b).

$U_g(x)$ as defined by eq. (B2-17) is a representation of the time
independent gauge group in the sense that $U_{g_1} U_{g_1} =
U_{g_1g_2}$. For an infinitesimal transformation

\noindent
$g(x) = 1 +
\varepsilon^a(x)T_a$ $(a = 1, \ldots, dimG)$ eq. (B2-17) gives:
$$
U_{1+\varepsilon^a\tau_a}\Psi(\phi(x)) = (1 +
\int\varepsilon^a(x)G_a(x)dx)\Psi(\phi(x)).\leqno{(B2-18)}
$$
Hence $G^a(x)$ are the generators of infinitesimal gauge transformations
and eq. (B2-6) is equivalent to the requirement
$$
U_g \Psi(\phi(x)) = \Psi(\phi(x))\leqno{(B2-19)}
$$
that is, by eq. (B2-17):
$$
\Psi(g^{-1}\phi g) = exp(i Tr\int(gdg^{-1}
\phi))\Psi(\phi)\leqno{(B2-20)}
$$
for every $g(x)$ continuously connected to the identity.

In particular, for a configuration of the type of eq. (B2-7) one obtains:
$$
\Psi(\phi(x) = h^{-1}(x)\phi_D h(x)) = exp(i
Tr\int(hdh^{-1}\phi_D))\Psi(\phi_D).\leqno{(B2-21)}
$$
This $\Psi(\phi(x))$ fulfills eq. (B2-6).

At this stage it seems that the only constraint on the function
$\Psi(\phi_D)$ is that it be normalizable in some appropriate measure. We
will proceed to find additional constraints on $\Psi(\phi_D)$.

The solution as expressed in eq. (B2-21) is less explicit than that of
eq. (B2-14), it requires to find the appropriate $h(x)$ for each
$\phi(x)$. In fact $h(x)$ is defined by $\phi(x)$ only up to a left
multiplication by an element of $T\times W$, where $T$ is the subgroup
generated by the Cartan subalgebra $C$ and $W$ is the Weyl group. If
$\phi_D$ is fixed to be in a particular Weyl chamber there is left only
an arbitrarity in $T \colon h(x)$ and $t(x)h(x)$ with $t(x)\varepsilon T$
correspond to the same $\phi(x)$. Therefore the phase factor in eq. (B2-21)
should
be invariant under $h \rightarrow th$.

Comparing the phases of eq. (B2-21) in both cases:
$$
Tr(thd(th)^{-1}\phi_D) = Tr(hdh^{-1}\phi_D) +
Tr(tdt^{-1}\phi_D)\leqno{(B2-22)}
$$
$t(x)$ can be written as
$$
t(x) = exp(\sum_{i=1}^r \rho^i(x)T_i)\leqno{(B2-23)}
$$
thus invariance would result from eqs. (B2-22) and (B2-23) if
$$
exp(iTr\int tdt^{-1}\phi_D) = exp(i \sum_{i} \int \partial \rho^i Tr(T_i
\phi_D)dx)\leqno{(B2-24)}
$$
would be equal to the unity matrix. As $\phi_D$ is constant this would
occur in particular for any periodic function $\rho^i(x)$, for any value
of the constant $\phi_D$.

However, the function $\rho^i$ appearing in the parametrization of eq.
(B2-23) need not be periodic. All that is required is that $t(x)$ be a
periodic solution.

For example, for a compact simply connected group, $t(x)$ is continuous
whenever
$$
\rho^i(2\pi R) = \rho^i(0) + \alpha^i\leqno{(B2-25)}
$$
where the $r$ dimensional vector $\alpha$ belongs to the dual of the
weight lattice of $G$ and $R$ is the radius of the spatial world sheet
circle. (For non compact groups, restrictions hold,
although there are not always lattices present).
Therefore, for $\Psi(\phi)$ to be single valued under two different
representations
of the same $\phi(x)$ in terms of $h$ and in terms of $th$, $\phi_D$ has
to belong to the weight lattice of $G$, modulo $W$, i.e. it
corresponds to some highest weight. In particular, for $G = SU(2)$,
for $\phi_D$, which is of the form $\phi_D = {{iM}\over 2}\tau_3$ $M$ has to be
an integer. For that group $\Psi(\phi)$ is restricted
to configurations for which
$$
\phi^2_1 + \phi_2^2 + \phi^2_3 = M^2, \leqno{(B2-26)}
$$
$M$ is an integer.

We will return later to consider the target space picture this implies.
We have obtained however the support of $\Psi(\phi)$ for simply
connected Lie groups $G$. If $G$ is not simply connected, for example if
$G = SO(3)$, then $t(x)$ would be considered periodic even if $t(0) =
-t(2\pi R)$. In that case $\alpha$ is shortened by one half and $M$
would thus have to be an even integer to allow for continuity. This
corresponds to $SO(3)$ having only integer angular momentum
representations. The
general observations is that, as anticipated, to each representation of
$G$ corresponds one state in the Hilbert space. For an allowed
configuration $\phi(x) = h(x)\phi_Dh^{-1}(x)$ the dependence on the
angular variables $h(x)$ is fixed by the phase as appearing in eq.
(B2-21). To appreciate the geometrical significance of the phase, we
recast it in a different form.

The expression (B2-21) is not, as we have discussed, explicit in terms
of $\phi(x)$. Moreover, we have found that locally there is a dependence
on the particular choice of $h(x)$ made. Restrictions on the allowed
values of $\phi_D$ ensure that at the end (that is after completing the
spatial integration in eq. (B2-21)), there is no dependence on the
representative of $h(x)$, and that is all that really counts. One can try
to adopt a particular arbitrary choice of $h(x)$ so as to free
$\Psi(\phi)$ from the ambiguity, in general, however such a choice can't
be made globally on $G/T$ without meeting singularities of
$\phi(x)$ as long as $G$ is a nontrivial bundle on $G/T$. There is an
alternative, $WZNW$ like, ambiguity free, formulation in terms of
$G/T$ only, with the price of the need to continue $\phi(x)$,
letting it take values in $G/T$ not only on the physical circle,
$S^1$, but on a two dimensional disc, $D$, having $S^1$ as a boundary.
Consider the phase appearing in eq. (B2-21)
$$
\int_{S^1}Tr(hdh^{-1}\phi_D) = \int_{D(\partial D = S^1)}d
(Tr(hdh^{-1}\phi_D) = -\int_D Tr(hdh^{-1}\wedge
hdh^{-1}\phi_D).\leqno{(B2-27)}
$$
Comparing the two form in eq. (B2-27) between two choices of $h(x)$,
$h(x)$ and $t(x)h(x)$, $t\varepsilon T\times W$, we have, using the
commutativity of $tdt^{-1}$ and $\phi_D$,
$$
\eqalign{
&Tr((thd(h^{-1}t^{-1}\wedge thd(h^{-1}t^{-1}))\phi_D) =
Tr(hdh^{-1}\wedge hdh^{-1}\phi_D)\cr
&+ Tr(tdt^{-1}\wedge t(hdh^{-1})t^{-1}\phi_D) +
Tr(t(hdh^{-1})t^{-1}\phi_D\wedge tdt^{-1}) \cr
&+ Tr((tdt^{-1})\wedge(tdt^{-1})\phi_d) = Tr(hdh^{-1}\wedge hdh^{-1}\phi_D).\cr
}\leqno{(B2-28)}
$$
Demonstrating that the two form expression for the phase is locally
gauge invariant under $h \to th$. Also, the contraction of this form
with a vector field in the fibre direction $\delta h = \tau h \quad \tau
\varepsilon C$, is, since $\delta h^{-1} = - h^{-1}\tau$
$$
Tr(h\delta h^{-1} hdh^{-1}\phi_D) - Tr(hdh^{-1}h\delta h^{-1}\phi_D) =
Tr(\tau hdh^{-1}\phi_D - hdh^{-1} \tau \phi_D) = 0.\leqno{(B2-29)}
$$
It is, therefore, actually a closed two form on $G/T$ (on $G$ it is
an exact form by its construction, not so in general, on $G/T$).
The wave functional can now be written as:
$$
\Psi(\phi(x) = h^{-1}\phi_Dh) = exp(-i\int_{D(\partial D =
L)}Tr(hdh^{-1}\wedge hdh^{-1}\phi_D)\Psi(\phi_D)\leqno{(B2-30)}
$$
where this time the phase depends only on the values of $\phi(x)$ as
continued into the disc.

In direct notation this phase is given by $f^{ijk}\phi_i d\phi_j \wedge
d\phi_k/Tr\phi^2$. The quantization of $\phi_D$ can also be addressed in this
notation. For example, in the compact case it results from the fact that
$\Pi_2(G/T) = Z^r$. There are many homotopically inequivalent
ways to choose a disc with a given boundary and the compatibility of
$\Psi$ among all these choices requires the quantization of $\phi_D$ on
the weight lattice of $G$. If $G/_T$ is not simply connected, then in
general $\phi(x)$ can't be continued into a disc having $L$ as a
boundary. Still, in a given homotopy sector one can choose some
reference configuration $\widehat\phi(x)$. Any other configuration
$\phi(x)$ in the same sector can be continued into a two dimensional
annulus, $A$, whose boundary is the union of the paths $\phi(x)$ and
$\widehat\phi(x)$ with opposite orientations. In that case eq. (B2-30)
should be replaced by:
$$
\Psi(\phi(x)) = [exp(-i\int_ATr(hdh^{-1}\wedge
hdh^{-1}\phi_D))]\Psi(\widehat\phi)\leqno{(B2-31)}
$$
where $h$ is still defined by eq. (B2-7). Changing the choice
$\widehat\phi(x)$ merely multiplies $\Psi(\phi)$ by a total phase.

Let us return now to the example, $G = SU(2)$.

We have already found the wave functional in that case (eq. (B2-14)),
let us now relate it to eq. (B2-30).

The scalar field $\phi(x)$ is parametrized as:
$$
\phi(x) = {i \over 2}(\phi_1(x)\tau_1 + \phi_2(x)\tau_2 +
\phi_3(x)\tau_3)\leqno{(B2-32)}
$$
$\tau_a$ being the standard Pauli matrices. Eq. (B2-6) has the form
$$
(\partial \phi^a - \varepsilon_{abc}\phi^a {\delta \over{\delta
\phi^i(x)}})\Psi(\phi(x)) = 0\leqno{(B2-33)}
$$
we have shown that
$$
\phi^2_1(x) + \phi^2_2(x) + \phi^2_3(x) = M^2
$$
is $x$ independent.

For such configurations $\phi$ varies on ${G \over T} = {{SU(2)}\over{U(1)}} =
S^2$.

 $\phi(x)$ for which
$\Psi(\phi(x))$ has support is written as:
$$
\phi(x) = {i \over 2}(M \sin \theta(x)\cos \varphi(x)\tau_1 + M \sin
\theta(x)\sin \varphi(x)\tau_2 + M \cos \theta(x)\tau_3).\leqno{(B2-34)}
$$
The group element $h(x)$ which conjugates $\phi$ to $\phi_D$ is:
$$
h(x) = exp(i\varphi \tau_3)exp(- i {{\theta(x)}\over 2}(\sin \varphi(x)
\tau_1 - \cos \varphi(x)\tau_2))\leqno{(B2-35)}
$$
$$
h\phi h^{-1} \equiv \phi_D = {i \over 2} M\tau_3 \varepsilon
C\leqno{(B2-36)}
$$
for any function $\varphi(x)$. Making the arbitrary choice $\varphi = 0$, we
get

for the phase in eq. (B2-21)
$$
Tr(hdh^{-1}\phi_D) = -{M\over 2}(1 - \cos \theta)d\phi\leqno{(B2-37)}
$$
so:
$$
\Psi(\theta(x), \phi(x)) = exp(-i {M\over 2} \int(1 - \cos
\theta)d\phi)\Psi({i\over 2}M\tau_3).\leqno{(B2-38)}
$$
The same result follows from eq. (B2-14) using eq. (B2-34).

To be more precise this occurs after adding to the phase in eq. (B2-14)
a total derivative $M \int d\phi$. This freedom is fixed by requiring
continuity near $\theta = 0$ (the north pole) when a loop that does not
contain inside it the north pole is smoothly deformed to contain it.
$\theta, \phi$ have indeed two points of singularity on the sphere.
Near the south pole it is the quantization of $M$ that ensures the
continuity. The exterior derivative of the 1 form $(1 - \cos
\theta)d\phi$ is given by:
$$
d((1 - \cos \theta)d\phi) = \sin \theta d\theta \wedge d
\phi\leqno{(B2-39)}
$$
which is just the volume form on the sphere. In the version of eq.
(B2-32) the wave functional is thus:
$$
\Psi(\phi(x)) = exp(- i {M\over 2} \, (\rm{area \, enclosed
\, on \, the \, unit \,
sphere}\,
S^2 \, \rm{by \, the \, curve} \, \phi(x))) \Psi({i\over
2}M\tau_3).\leqno{(B2-40)}
$$
There are many ways to define a disc on $S^2$ with a fixed boundary.
Their areas differ from each other by integer multiplies of the total
area of the sphere. We thus reobtain the quantization condition
$$
{M\over 2}4\pi = 2\pi \quad \rm{integer} \quad \Rightarrow M\varepsilon
Z.\leqno{(B2-41)}
$$
Therefore, in general, the phase is just the appropriate area bounded by the
configuration $\phi(x)$.

In the 3-d Chern-Simons case [9], [10] the phase of the wave
functional itself correspond to a 2-d action, that of the chiral
WZNW. In the present case the phase is again
an action, this time of a quantum mechanical system [11] containing also
only first order derivatives.
\v
\noindent
{\bf B3} $\phi \cdot F$ {\it theory for some non-compact groups.}
\v
The above analysis holds also for non-compact groups.

In particular we will consider $G = SO(2, 1), SO(2, 1) \times U(1)$, a
contracted version of $SO(2, 1) \times U(1)$ and $SO(2, 2)$.

These theories are related, in some sense, classically to various two
dimensional gravitational systems. This relationship will be discussed
in greater detail in section C.

Starting with $G = SO(2, 1)$, one notes
that $G$ has three types of adjoint orbits, the generators being
elliptic, hyperbolic and parabolic. Denoting $\phi(x)$ by
$$
\phi(x) = i\phi_1 \tau_1 + i\phi_2 \tau_2 + \phi_3
\tau_3.\leqno{(B3-1)}
$$
A configuration $\phi(x)$ can be brought to the form $\phi_D$ of eq.
(B2-7) depending on the sign of $M^2$, where now:
$$
{1 \over 2} tr\vec\phi^2 = \phi^2_3 - \phi^2_1 - \phi^2_2 =  M^2\leqno{(B3-2)}
$$
$M^2$ remains $x$ independent as we have shown. For a positive value of
$M^2$, $\phi(x)$ can be brought to the form $M\tau_3$ ($M$ real). The
orbit given by eq. (B3-2) represents a two sheeted hyperboloid, as the
stability groups of such orbits are compact for $M^2 > 0$, the analysis in
the compact case $G = SU(2)$ applies here as well, rendering $M$ to be
an integer. As $SO(2, 1)$ contains no operation relating the two disjoint
parts of the two sheeted hyperboloid, there exist two different classes
of representations denoted by $M$ positive or negative. They are in one
to one correspondence to the discrete series representations, $D_+,
D_-$, if $SO(2, 1)$. Representations for which the invariant Casimir is
negative and the eigenvalue of the Cartan subalgebra generator is
bounded from either above or below.

Negative values of $M^2$, lead to orbits on the one sheeted hyperboloid, in
which case the generator of the stability group is a hyperbolic, non
compact operator. Thus $\phi(x)$ can be brought to the form
$iM\tau_1$, no quantization is needed to ensure that $t = exp(\alpha
\tau_1)$ be continuous. To every real value of $M$ corresponds a basis
element of the wave functional. The states are in one to one
correspondence to the so-called principal series representations of
$SO(2, 1)$, which have no restriction on the value of the real Casimir. The
orbits are classified by a fixed radius of the one sheeted hyperboloid.

For $M^2 = 0$ $\phi$ can be brought to the form $\phi = i\phi_1\tau_1
+ \phi\tau_3$ and one distinguishes two cases. The first orbit is
$\vec\phi = 0$, which is represented by the tip of the light one,
the second orbit is the light cone itself. The first is related to the
singlet representation of $SO(2, 1)$, we have not found an orbit associated
with the complementary
representations of $SO(2, 1)$.

The area phase, for positive $M^2$, can be chosen to be the finite area
contained by $\phi(x)$. For negative values of $M^2$, the cases can be
classified according to their winding number around the axis of the
hyperboloid.
All  the different sectors are still gauge equivalent to each other
by topologically non-trivial gauge transformations of the non-simply
connected gauge group.
For any non zero winding, one needs to measure the area
relative to some standard curve. For zero winding, there again exists a
finite area which can be chosen.  Since the gauge transformations which lead
from one winding sector to another are not continuously connected to the
identity the vanishing of the infinitesimal generators $G^a$, i.e. Gauss'
law, does not fix the relative values of the wave function at different
sectors.   If we insist on a wave functional invariant under finite
gauge transformations defined by eq. (B2-17), then the value of the
wave functional at zero winding configurations fixes its value on all the
other sectors.  In this non-simply connected case there exists however,
the possibility of a consistent modification of the transformation law eq.
(B2-17) by introducing a $\theta$ parameter multiplying eq. (B2-17) by $e^{in
\theta}$,
n being the winding number of the gauge transformation.  This will introduce
a relative factor $e^{in\theta}$ in the winding number n sector of the
wave functional.  We do not know though how to express such a theta parameter
in terms of a gauge invariant local action.

As  for $SU(2)$, one can obtain the explicit solution of Eq. (B2-6) in
terms of $\phi_1, \phi_2, \phi_3$. It is with the definitions of eq.
(B2-14)
$$
\Psi(\rho, \sigma, \alpha) = exp(\oint {{\rho\partial\sigma -
\sigma\partial\rho}\over{M^2 + \alpha^2}} \alpha dx)F(M^2).\leqno{(B3-3)}
$$
Again to the phase one has to add a part $M\oint dx$ for $M^2 > 0$. For
configurations with other values of $M^2$ it is redundant.

Let us next consider the group $G = SO(2, 1)\times U(1)$, in that case it
is enough to treat the $U(1)$ case, as the problem factorizes and the
$SO(2, 1)$ has just been studied.

For $U(1)$ the solution of eq. (B2-6) is immediate, the field $\phi_4$
associated to the $U(1)$ generator must be $x$ independent. If the
$U(1)$ gauge group is non compact, $\phi_4$ may assume any value, for a
compact $U(1)$, $\phi_4$ will be appropriately quantized in units of the
$U(1)$ radius.

Finally let us consider the twisted $SO(2, 1)\times U(1)$ algebra [8].

One can supplement the $SO(2, 1)$ algebra
$$
\eqalign{
&[J_0, P_\pm] = \pm P_\pm\cr
&[P_+, P_-] = J_0\cr
}\leqno{(B3-4)}
$$
by another generator $I$ of a $U(1)$ group making it an $SO(2, 1) \times
U(1)$ algebra, enabling a study of a twisted algebra of these four
generators
$$
\eqalign{
&[J_0, P_\pm] = \pm P_\pm\cr
&[P_+, P_-] = \lambda I.\cr
}\leqno{(B3-5)}
$$
$I$ commutes with all generators. This is still a Lie algebra which may
actually be reached by a continuous deformation of the $SO(2, 1) \times
U(1)$ case. The two operators $C_1 = \lambda I$, $C_2 = P_+ P_- +
\lambda I J_0$ are Casimir operators commuting with the algebra.
Consider again the action $S = T r\int
\phi F$ built on that algebra.
Here $F$ is the curvature of the gauge field based on the above algebra
and $\phi$ is a four component Higgs field in the adjoint
representation:
$$
\phi = \phi^+ P_+ + \phi^- P_- + \phi^0 J_0 +
\phi^II.\leqno{(B3-6)}
$$
The symbol $Tr$ denotes here the invariant bilinear form of the algebra
[8]
$$
Tr(\phi F) = \phi^+ F^- + \phi^- F^+ + {1 \over
\lambda}(\phi^I F^0 + \phi^0 F^I).\leqno{(B3-7)}
$$
Under a gauge transformation $\phi^0$ is unchanged and the combination
$\phi^+ \phi^- + {1 \over \lambda} \phi^0 \phi^I$ is
also invariant. Gauss's law thus implies that the wave functional
obtains support only on configurations for which
$$
\phi^+ \phi^- + {C \over \lambda}\phi^I =
M^2\leqno{(B3-8)}
$$
$C$ and $M^2$ (the defining radius of the parabolic hyperboloid) are
constant in space. An invariant metric on our Lie algebra is
$$
ds^2 = a^2(d\phi^+ d\phi^- + {1 \over \lambda}d\phi^0
d\phi^I) + b^2(d\phi^0)^2\leqno{(B3-9)}
$$
$a$ and $b$ are arbitrary constants. On a given orbit the metric will be
just $a^2d\phi^+ d\phi^-$. Every $\phi(x)$ configuration on
such an orbit can be brought by a gauge transformation $h(x)$ to the $x$
independent form: $h(x) \phi(x)h^{-1}(x) = \phi$ with, say,
$\phi_D^\pm = 0$, $\phi^0_D = C$, $\phi_D^I = {{\lambda
M^2}\over C}$. The transformation $h$ is defined modulo right
multiplication by a transformation $t$ which preserves $\phi_0$ i.e.
$t$ is generated by $J_0$ and $I$ with no $P_\pm$. The corresponding
phase in the wave functional $Tr \int(hdh^{-1}\phi_D)$ is ambiguous
by a phase
$$
Tr\int(tdt^{-1}\phi_0) = Tr\int(d\alpha^0J_0 +
d\alpha^II)(CJ_0 +{{\lambda M^2}\over C}I) = \int {{\lambda^2
M^2}\over C} d\alpha^0 + \int \lambda Cd\alpha^I.
$$
Since $J_0$ generates a non compact group, the first term vanishes. If
however, $I$ generates a compact $U(1)$, then the second term may give
$\int d\alpha^I =$ integer which imposes quantization on possible values
of $\lambda \phi^0 = \lambda C$.
\v
\noindent
{\bf B4} {\it Some Gauge theory aspects.}
\v
The $\phi \cdot F$ theory is a gauge theory, as it is also a topological
theory all of the states in the theory have quantum mechanically zero
energy thus falling in the category of flat potentials. For all groups $G$
the symmetry group remains $G$ if the
system is chosen to be in the state corresponding to the singlet. (For
example for $SU(2)$ this implies: $\Psi(\phi) = \delta(\phi)$).
A generic result is that for a wave function supported on any
other orbit the symmetry is reduced at most to $\displaystyle\prod_{i=1}^r
H_i$. $H_i$
are various one dimensional groups ($U(1)$'s for compact groups) and $r$
is the rank of the group. Indeed, for compact gauge groups, a scalar as
$\phi$ field in the adjoint representation reduces $G$ to a product of
groups whose ranks add up to the rank of $G$, generically reducing $G$ to
$(U(1))^r$. For a flat potential $G$ can either be maintained or broken down to
the above factors.

The wave function supported on a fixed orbit allows $\phi(x)$ in general
to vary in space. The identification of the residual $H_i$ symmetries is
more complex. For example one may employ the methods used for the
monopole configuration [12]. This will be further
discussed in relation to gravity.

Treating $\phi \cdot F$ as a gauge theory one may calculate it's order
parameters. We have already demonstrated that the operator $O$, defined
by
$$
O = tr \phi^2(x)\leqno{(B4-1)}
$$
is $x$ independent (as it should be in a topological theory) and it's
expectation value is given by
$$
<0> = \int |\Psi(\phi_D)|^2 (tr \phi^2_D)D\phi_D.\leqno{(B4-2)}
$$
Note that the phase of $\Psi(\phi(x))$ and with it all $h(x)$ dependence
is eliminated. For $\Psi(\phi_D)$ on a fixed orbit $\phi^f_D$
$$
<0> = tr(\phi^f_D)^2.\leqno{(B4-3)}
$$
Another gauge theory order parameter is the value of the Wilson loop. In
the Hamiltonian formulation the calculation of the average value of the
Wilson loop reduces to an addition of two sources to the system.

This is done in the following manner:

Suppose the expectation value of a Wilson loop in the representation $R$
is inserted. One deals then with the path integral:
$$
Z(A) = \int D\phi DA_0DA e^{Tr\phi F}
Tr_RPe^{\int_C A}.\leqno{(B4-4)}
$$
Choose the Wilson line $C$ to run along the $t$ axis at $x = 0$.
Consider, first the Abelian case. The representation $R$ is fixed then
by the charge $q$ of the loop and one computes the expectation of
$exp(q \int A)$. In the gauge $A_0 = 0$, varying with respect to $A_0$ Gauss'
 law
is changed into
$$
[d\phi + q\delta(x)] = 0.\leqno{(B4-5)}
$$
Naturally the flux of the charged particle enters into Gauss' law and
the wave function is supported on configurations of $\phi$ which is
constant everywhere except at the position of the particle where it
makes a jump of height $q$ which represents the flux of the particle. If
space is a compact circle, it must contain zero total charge and another
particle of charge $-q$ has to be added, say, at the point $x = L$. The
allowed configurations are then those with a constant $\phi = M$
everywhere except for the interval between the two opposite charges
where the flux line between them is manifested as a shift in $\phi$
to the value $M - q$. Notice however that in this topological theory the
flux line carries no energy density and does not really confine the
sources.

In the non abelian case the external source, although not moving in
space, still carries a discrete dynamical degree of freedom which is
its orientation in the gauge group. The wave functional
$\Psi_s[\phi(x)]$ depends now, for a single source, also on the index
$s$,

\n
$dim(R) \geq s \geq 1$, representing the gauge state of the source
Gauss' law will now obey
$$
\sum_{u = 1}^{dim(R)}[(D\phi)^a \delta_{s, u} + \delta(x)\lambda^a_{s,
u}]\Psi_u[\phi(x)] = 0\leqno{(B4-6)}
$$
for every point $x$, every group index $a$ and every $R$ representation
index $s$. If $\phi(x)$ is conjugate to $\phi_D(x)$ which is in the
Cartain subalgebra $C$: $\phi_D(x) = h(x) \phi(x)h^{-1}(x)$

\n
$\phi_D(x)\in
C$ then the same reasoning as we had in the absence of sources states
that Gauss' law implies
$$
\sum_{u = 1}^{dim(R)}[\partial \phi^i_D(x)\delta_{s, u} +
\delta(x)\lambda^i_{s, u}] \Psi_u = 0\leqno{(B4-7)}
$$
for every $i$ in $C$, $r \geq i \geq 1$ $r$ being the rank of $G$.
Choose the basis in the representation $R$ to be the weight basis with
respect to the Cartan algebra $C$ i.e. for any $r \geq i \geq 1$
$$
\sum_{u = 1}^{dim(R)} \lambda_{s, u}^i \Psi_u =
w^i_s\Psi_s\leqno{(B4-8)}
$$
with $w_s$ the weight corresponding to the $s$-th state in the $R$
representation. Gauss' law implies then that our wave function may
differ from zero only on configurations of the form
$$
\phi_{D, s}^i(x) = \cases{m^i & $x < 0$\cr
m^i - w^i_s & $x > 0$\cr}\leqno{(B4-9)}
$$
with constant $m^i$ and $w_s$ being one of the weights of $R$.
At the configuration $\phi(x) = \phi_{D, s}(x)$ only the components of
$\Psi$ corresponding to $w_s$ are non zero. We can normalize our wave
function by
$$
\Psi^{(s)}_r (\phi(x)) = \phi_{D, s}(x)) = \delta_{r, s}.\leqno{(B4-10)}
$$
(If the weight $w_s$ has multiplicity $m$, we can define $m$ different
wave functions by
$$
\Psi_{r, K}^{(s, \ell)}(\phi(x) = \phi_{D, s}(x)) = \delta_{r,
s}\delta_{K, \ell}\leqno{(B4-11)}
$$
$m \geq \ell \geq 1$).

To find the value of $\Psi^{(s)}$ at a general admissible $\phi$
configuration of the form
$$
\phi(x) = h^{-1}(x)\phi_{D, s}h(x)\leqno{(B4-12)}
$$
we have only to note that in the presence of a source the gauge
transformation of the wave function should be:
$$
(U_{g(x)}\Psi)_s[\phi(x)] = \sum_{\tau = 1}^{dim R}D(g(0))_{s,
r}e^{-iTr\int (gdg^{-1}\phi)}\Psi_r(g^{-1}(x)\phi(x)g(x))
\leqno{(B4-13)}
$$
where $D^{(R)}(g)$ is the finite matrix representing $g$ in
representation $R$. Again, Gauss' law forces $\Psi$ to be invariant under
$U$, hence
$$
\Psi_r[h^{-1}(x)\phi_{D, s}h(x)] = D^{(R)}(h^{-1}(0))_{r,
s}e^{iTr\int (h(x)dh^{-1}(x)\phi_{D, s})}.\leqno{(B4-14)}
$$
There is a state $\Psi$ for every vector in the representation $R$.
Again, for a compact circular space the configuration $\phi_{D, s}$ is
not legitimate for $s$ a non zero weight. Then there must be at least
one other source, say at the point $x = L$, in a representation $R'$
containing the weight $-w_s$. The wave function depends now on the
indices of both sources. At the Cartan configuration
$$
\phi^i_{D, s, -s}(x) = \cases {m^i - w_s^i & $0 \le x \le L$ \cr
m^i & otherwise \cr}\leqno{(B4-15)}
$$
we have by Gauss' law
$$
\Psi^{(s)}_{r, r'}[\phi(x) = \phi^i_{D, s, -s}] = \delta_{r,
s}\delta_{r', -s}.\leqno{(B4-16)}
$$
If $m$ and $m'$ are the multiplicities of $s$ in $R$ and $R'$
respectively, then we have $mm'$ independent wave functions and for a
general admissible configuration
$$
\Psi^{(s)}_{r, r'}[h^{-1}\phi_{D, s, -s}h(x)] =
D^{(R)}(h^{-1}(0))_{r, s}D^{(R')}(h^{-1}(L))_{r',
-s}e^{iTr\int hdh^{-1}\phi_{D, s, -s}}.\leqno{(B4-17)}
$$
Note that if $R$ contains the zero weight, then a single source is
possible even in compact space. This is somewhat analogous to the
solution in $QCD$ where zero colour triality sources are not confined but
rather screened by gluons.

Actually recall that the theory described by:
$$
S = \int_\Sigma d^2xTr(\phi \cdot F + e^2\sqrt g \phi^2)
\leqno{(B4-18)}
$$
would give pure $QCD$ in two dimensions once the $\phi$ degrees of
freedom are integrated upon to yield
$$
S = \int_\Sigma d^2x\sqrt g {1 \over {e^2}}TrF^2\leqno{(B4-19)}
$$
$e^2$ being a dimensionfull coupling.

The Hilbert space of the action given by eq. (B4-18) is identical to the
Hilbert space described the action of eq. (B1-1) in the $A^i_0 = 0$
gauge. This follows as the additional term $e^2 tr \phi^2$ is $A_0$
independent. It is however metric dependent and thus modifies the energy
of the system in a way similar to a magnetic field in the Zeeman-effect.
 As $tr\phi^2$ is constant the action depends only on the area. It is
slightly disappointing that a theory with seemingly less symmetry such
as pure $(QCD)_2$ is so similar to the topological theory. For any group $G$,
the infinite degeneracy is lifted by the
$e^2tr\phi^2$ term.

The exact value of the energy can be
renormalized, one expects, from pure $(QCD)_2$, that the renormalized
energy is this case is not $tr\phi^2$, but the second Casimir. In any
case the wave functionals themself are left unchanged.

For the $\phi \cdot F$ theory we have found that the wave functionals
behave as in strongly coupled lattice $QCD$, in particular $QCD$ is
manifestly confining in that region of parameter space.

The formation of the flux lines, in the case when the inserted
representation is sensitive to the group center, is the clear geometrical
(stringy) feature of confinement.

The next step in any such evaluation is to measure the dependence of the
energy difference between the energy of the system with inserted charges
and without such charges on the distance between the sources. For
confining systems it grows linearly with that distance. In particular
when a chromo-electric flux line is formed among the sources confinement
is manifest.

For the topological theory the energy of all states, with or without
sources remains zero (if the sources can be coupled in a topological
manner, which is possible in the $\phi \cdot F$ case) and thus no
dependence of the distance between the sources can emerge. Indeed any
such dependence would violate the topological nature. However the moment
the $tr \phi^2$ term will be switched on, the flux line will have a fixed
energy per unit length and the wave functionals in the
presence of sources (which remain unchanged when $e^2tr\phi^2$ was turned
on) will now indeed signify that the confinement properties of pure
$(QCD)_2$.

The scalar field $\phi$ playing the role of a chromoelectric field,
whose constancy in $(QCD)_2$ is well known.
\v
{\bf B5} {\it The case of a space with a boundary.}
\v
For a closed spacial space each representation is one to one
correspondence with each state in the Hilbert space.

Consider now the  model on a space time manifold with a
boundary. Let us adopt boundary conditions which force the component of
the gauge field tangent to the boundary to vanish. Note that with this
condition there is no boundary correction to the bulk equations of motion
[10]. In the Hamiltonian picture we have a space which
is an interval, say $[0, 1]$, evolving in time, with $A_0 = 0$ on the
boundaries. Since Gauss' law results from integration over $A_0$, it
does not have to be satisfied on the two edges of the spatial interval. As a
result our wave function has to be invariant only under those time
independent gauge transformations $h(x)$ for which $h = 1$ on the edges.

Choosing again the $\phi$ picture, this conditions is enough to ensure
that the wave functional $\Psi[\phi(x)]$ is non zero only for $\phi(x)$
which stays on the same gauge orbit throughout the interior of the
interval and hence, by continuity, also at the boundaries. It is still
true then, that admissible configurations are of the form, $\phi(x) =
h^{-1}(x)\phi_D h(x)$. Unlike the close string case, not any two
admissible configurations on the same orbit are connected by a gauge
transformation, since gauge transformations have to tend to unity on the
boundaries and hence $\phi(0)$ and $\phi(1)$ are gauge invariant. It is
still true that any two configurations with the same boundary $\phi$
values, are connected by a gauge transformation, and knowing the value
of the wave functional on one of them, Gauss' law determines its value on
the other. The most general physical wave functional will depend now on
an arbitrary function of the orbit and of $\phi(0)$ and $\phi(1)$. For
any pair of points $\phi(0)$ and $\phi(1)$ in the Lie algebra, belonging
to the same orbit fixed by some point $\phi_D$ in the Cartan subalgebra,
choose some arbitrary path $\overline\phi(x)$ in the same orbit
connecting these two points, i.e. $\overline\phi(0) = \phi(0)$, and
$\overline\phi(1) = \phi(1)$. The path $\overline\phi(x)$ is arbitrarily
fixed for each pair, for example it may be chosen as the geodesic line
on the $G/T$ orbit manifold connecting $\phi(0)$ to $\phi(1)$. The value
of the wave functional at these configurations $\overline\phi(x)$ is not
restricted by gauge invariance and may be arbitrarily put to be any
function of the orbit $\phi_D$, and the two boundary values which
determine $\overline\phi(x)$,
$$
\Psi[\overline\phi(x)] = f(\phi_D, \phi(0), \phi(1)).\leqno{(B5-1)}
$$
The value of $\Psi$ at any other configuration is now fixed by gauge
invariance. If $\phi(x)$ is any configuration in the orbit of $\phi_D$,
then there exists a transformation $h(x)$ such that, $\phi(x) =
h^{-1}(x)\phi_D h(x)$, which, however does not tend to unity at the
edges. For the $\overline\phi$ trajectory belonging to the same orbit
$\phi_D$ with the same boundary values $\phi(0)$ and $\phi(1)$, let
$\overline h(x)$ be a transformation such that, $\overline\phi(x) =
\overline h^{-1}(x)\phi_D \overline h(x), \overline h(x)$ also does not
tend to unity at the boundary. Since at the edges $\phi$ and $\overline\phi$
agree, we can choose $h(0) = \overline h(0)$ and $h(1) = \overline
h(1)$. The configurations $\phi(x)$ and $\overline \phi(x)$ are
connected by the gauge transformation,
$$
\phi(x) = k^{-1}(x)\overline \phi(x)k(x)\leqno{(B5-2)}
$$
with $k(x) = \overline h^{-1}(x)h(x)$, being a legitimate gauge
transformation with the right boundary behaviour. Due to the invariance
of $\Psi$ under such gauge transformations, the value of $\Psi$ at the
configuration $\phi(x)$ is given by
$$
\Psi[\phi(x)] = e^{i\int Tr(kdk^{-1}\overline\phi(x))}f(\phi_D, \phi(0),
\phi(1))\leqno{(B5-3)}
$$
or
$$
\Psi[\phi(x)] = e^{i\int [Tr(hdh^{-1}\phi_D) - Tr(\overline h d\overline
h^{-1}\phi_D)]}f(\phi_D, \phi(0), \phi(1))\leqno{(B5-4)}
$$
which completely determines $\Psi$ up to the arbitrary function $f$. The
phase above can further be expressed as an integral over a closed path
parametrized by $2 > y > 0$, $\int Tr[j(y)dj^{-1}(y)\phi_D]$, where
$j(y)\in G$ is defined by
$$
j(y) = h(y) \qquad \rm{for} \, 1 > y > 0\leqno{(B5-5)}
$$
$$
j(y) = \overline h(2 - y) \qquad \rm{for} \, 2 > y > 1.
$$
Again $j(y)$ is defined only modulo left multiplication by Cartan torus
elements $t(y)$ and the requirement of the independence of $\Psi$ under
such changes gives the same quantization of allowed orbits that we had
in the closed case. One can replace the phase in the above equation
which depends locally on $t(x)$, by the phase,
$$
e^{-i\int_D Tr(\ell(x, z)d\ell^{-1}(x, z) \wedge \ell(x,
z)d\ell^{-1}(x, z)\phi_D)}\leqno{(B5-6)}
$$
which is locally $t$ independent. Here $\ell(x, z)$ is a mapping from a
2 dimensional disc $D$, to $G$, which tends to $j(y)$ on the circle
bounding the disc.

In the $A^a$ basis (discussed in eq. (B2-4)) a basis may be formed out
of any function of the form
$$
\widetilde\Psi(A) = Tr Pexp \int_0^1 A^a_\mu dx_\mu.\leqno{(B5-7)}
$$
$\widetilde\Psi(A)$ is invariant under $\widetilde g(x)$
transformations. One is not restricted only to functions of the
characters of these group elements.

Let us now put forward a speculation on how a transition could occur
which would increase the number of states. One of the difficulties in
imagining a transition between a topological gravitational theory and a
non-topological phase of gravity is that a non topological phase of
gravity seems to consist of many more states. Imagine that a transition
on the world sheet would result in tearing the world sheet, in
particular in turning closed space to many open spaces. If this would
occur in a theory of the type $\phi \cdot F$, it would result in the
formation of many more states. Exactly how many more is determined by
the general nature of the group (compact or non compact) and on its
exact specification. Is there a mechanism which would allow to tear the
world sheet. It was pointed out [13] that one aspect of
crossing the $c > 1$ barrier is the rapture of the world sheet. Similar
phenomena occur at the K.T. transition. Above $c = 1$ the 2-d
gravitational system changes from a zero number of local degrees of
freedom system
to one with a positive number of such degrees of freedom.

\vs
\noindent
{\bf C. Gravitational aspects for the gauge theory.}
\v
\n
{\bf C1.} {\it Review.}
\v
We have solved quantum mechanically the topological gauge theory, the
promise of an emergence of a metric actually appears classically. Thus,
we retreat to the classical description and later return to search for a
fulfillment of these classical promises in the quantum setting. In
particular we discuss classically the case of the non-compact gauge group
$SO(2, 1)$. The theory has three generators which are denoted by $P_1,
P_2, J$ they fulfill the algebra:
$$\eqalign{
&[J, P_a] = 2i\varepsilon_a^bP_b \cr
&[P_a, P_b] = -i\Lambda \varepsilon_{ab}J.\cr}\leqno{(C1-1)}
$$
For now we suppress the scale $\Lambda$ and set $\Lambda = 1$.

As the theory has three gauge generators, it was suggested [4] to
relate them to the parameters associated with the two dimensional
diffeomorphisms and the local Lorentz tranformation similarly  to what
was
proposed earlier for a 3-d case [2]. This was shown [4] in
the following manner.

Under a gauge transformation $\lambda = \lambda^i J_i$ ($J_i$ denote
here the group generators), the gauge fields, $A_\mu = A^i_\mu J_i$, and
the scalar field, $\phi = \phi^i J_i$, transform infinitesimally respectively
as:
$$
\eqalign{
&\delta A_\mu = -\partial_\mu \lambda - [\lambda, A_\mu] \equiv - D_\mu
\lambda\cr
&\delta \phi = [\lambda, \phi]\cr
}\leqno{(C1-2)}
$$
under a diffeomorphism $\delta x^\mu = \varepsilon^\mu(x)$ the fields
transform as
$$
\eqalign{
&\delta A_\mu = \varepsilon^\rho F_{\rho \mu} + D_\mu(\varepsilon^\rho
A_\rho)\cr
&\delta \phi = \varepsilon^\rho D_\rho \phi - [\varepsilon^\rho
A_\rho, \phi].\cr
}\leqno{(C1-3)}
$$
As the equation of motions are
$$
F^{\mu \nu} = 0\leqno{(B1-2)}
$$
$$
D_\mu \phi = 0\leqno{(B1-3)}
$$
gauge invariance guarantees general coordinate transformations on
configurations which are solutions of the equations of motions; the
gauge and coordinate transformations are related by:
$$
\lambda^i = \varepsilon^\rho A^i_\rho.\leqno{(C1-4)}
$$
That is, general coordinate transformations define configuration
dependent gauge transformation (on the configuration $A^i_\rho = 0$, for
example, $\lambda^i = 0$). This may be implemented for any gauge group
$G$. Choosing two group directions ($i = 1, 2$) one may determine
$\varepsilon_\rho$ from $\lambda^i$ if $det_{\rho i} A^i_\rho \not= 0$.
We shall see later on how to reconstruct a general coordinate invariance
plus Lorentz invariance from the gauge group apart from the zero determinant
obstruction (topological defects), where the gauge transformation cannot
be inverted.

Moreover, it was suggested to define
$$
A_\mu = e_\mu^a P_a + w_\mu J\leqno{(C1-5)}
$$
such that $e^a_\mu, w_\mu$ play the role of zweibeins and spin connections
respectively. In general a metric $g_{\mu \nu}$ can be defined from the
zweibeins by:
$$
g_{\mu \nu} = e_\mu^a e^b_\nu B_{ab}\leqno{(C1-6)}
$$
where $B_{ab}$ is the $SO(2, 1)$ group metric restricted to the
transverse direction:
$$
\left(\matrix{1&0\cr
0&-1\cr}\right).\leqno{(C1-7)}
$$
Introducing $P_\pm = P_1 \pm P_2$, we find from eq. (C1-1)
$$
\eqalign{
&[J, P_+] = -2iP_+\cr
&[J, P_-] = 2iP_-\cr
&[P_+, P_-] = 2i J.\cr
}\leqno{(C1-8)}
$$
The field components are defined by
$$
\eqalign{
&2\phi = \rho P_+ + \sigma P_- + \alpha J\cr
&2A_\mu = e_\mu P_+ + \lambda_\mu P_- + \beta_\mu J\cr.
}\leqno{(C1-9)}
$$
Thus also:
$$
\eqalign{
2F_{\mu\nu} = &((\partial_\mu e_\nu - \partial_\nu e_\mu) - i\beta_\mu
e_\nu + i\beta_\nu e_\mu) P_+\cr
&+ ((\partial_\mu \lambda_\nu - \partial_\nu \lambda_\mu) + i\beta_\mu
\lambda_\nu - i\beta_\nu \lambda_\mu)P_-\cr
&((\partial_\mu \beta_\nu - \partial_\nu \beta_\mu) + ie_\mu \lambda_\nu -
ie_\nu \lambda_\mu)J = 0.\cr
}\leqno{(C1-10)}
$$
The equations of motions (B1-3) are explicitly
$$
\eqalign{
&- \partial_\mu \rho + i\rho\beta_\mu - i\alpha e_\mu = 0 \cr
&- \partial_\mu \sigma - i\sigma\beta_\mu + i\alpha \lambda_\mu = 0\cr
&\partial_\mu\alpha - i\rho\lambda_\mu + i\sigma e_\mu = 0.\cr
}\leqno{(C1-11)}
$$
The two equations requiring the vanishing of the $P_+$ and $P_-$
components of the electric field can be used to express the gauge field
$\beta_\mu$ in terms of the other two gauge fields $\lambda_\mu$ and $e_\mu$
by:
$$
\beta_\mu = {{e_\mu \varepsilon^{\rho\sigma}\partial_\rho \lambda_\sigma
+ \lambda_\mu\varepsilon^{\rho\sigma}\partial_\rho e_\sigma}\over{det(e,
\lambda)}} \qquad det(e, \lambda)\equiv
\varepsilon^{\mu\nu}e_\mu\lambda_\nu.\leqno{(C1-12)}
$$
Assigning to $e_\mu, \lambda_\mu$ the zweibein roles, the $J$ component in
eq. (C1-10) can be rewritten as:
$$
R = 1
$$
where $R$ is defined by analogy to be:
$$
R \equiv {{\varepsilon^{\mu \nu}\partial_\mu\beta_\nu}\over{det(e,
\lambda)}}.\leqno{(C1-13)}
$$
$R$ looks as a Ricci curvature scalar. The gauge equations of motion can
be reinterpreted as an equation for constant curvature which could have
been derived from a purely gravitational theory of the form
$$
S = \int \eta(R - 1)\sqrt gd^2x\leqno{(C1-14)}
$$
where $\eta$ is a Lagrange multiplier [14]. The reinterpretation
of the symmetries of the equations of motion and the possibility to
recast the gauge equations in a gravitational form suggest the emergence
of a metric in the topological gauge theory.

Before trying to understand what this classical metric emergence means
quantum mechanically we wish to reinstate a role for the scalar fields
$\phi^i$ in the analysis. First note that the equations of motion
(C1-11) require that
$$
tr\phi^2 \equiv M^2 = 2\rho \sigma - \alpha^2\leqno{(C1-15)}
$$
be constant in both space and time. In section B we have seen that the
spatial components of eq. (C1-11) were the Gauss's law in the $A_0 = 0$
gauge so that
$$
\partial_1(M^2) = 0\leqno{(C1-16)}
$$
was also valid at the quantum level. The vanishing of $\partial_t
M^2$ follows also quantum mechanically from the vanishing of
$H$.

\v
\n
{\bf C2.} {\it Constructing an $H$ invariant metric.}
\v
Following the gauge theory point of view, as expressed in the former
section, any classical choice of a configuration $\phi$ of a given
''length'' $M^2$, would result in the ''breaking'' of the group $G =
SO(2, 1)$ down to a one generator sub group $H$ of $SO(2, 1)$. As the
Hamiltonian vanishes identically, the classical ''potential'' is flat
and any configuration $\phi^a$ of given length $M^2$ is allowed. For
$M^2 > 0$ $SO(2, 1)$ could be viewed as broken to a compact $U(1)$,
while for $M^2 < 0$ the residual symmetry is a non compact $O(1, 1)$.

In both cases the axis of the remaining symmetry is determined by the
configuration $\phi^a$. (The $M^2 = 0$ cases will be treated separately).
Realizing that residual $SO(2, 1)$ gauge invariance will require a well
defined superposition of all $\phi^a$ on an orbit of a given allowed
$M^2$, we follow the original suggestion as used for the monopole
configuration [12].

The gauge fields are decomposed in terms of a longitudinal coordinate,
parallel to the $\phi^a$ field and two transverse coordinates in the
orthonormal directions.

For this purpose we choose the following orthogonal basis
$$
\eqalign{
&\widehat e_L = {\phi \over {|\phi|}} \equiv {\phi \over{\sqrt{M^2}}} =
{1 \over {\sqrt{M^2}}}[\rho P_+ + {{M^2 + \alpha^2}\over{2\rho}}P_- +
\alpha J]\cr
&\widehat e_{T_1} = {1 \over{\sqrt{-(M^2 + \alpha^2)}}}(-\rho P_+ + {{M^2
+ \alpha^2}\over{2\rho}}P_-)\cr
&\widehat e_{T_2} = \sqrt{{{-\alpha^2}\over{M^2(M^2 +
\alpha^2)}}}(\rho P_+ + {{M^2 + \alpha^2}\over{2\rho}}P_- + {{M^2 +
\alpha^2}\over \alpha}J)\cr
}\leqno{(C2-1)}
$$
here we have expressed the field $\sigma$ in terms of $\rho$ and
$\alpha$.

Decomposing in that basis the vector field $A_\mu$
$$
A_\mu = E_{L_\mu}\widehat e_L + E_{T_{1\mu}} \widehat e_{T_1} +
E^1_{T_{2\mu}}\widehat e_{T_2}\leqno{(C2-2)}
$$
$E_{L_\mu}$ is the gauge field component in the direction of the surviving
symmetry.  The transition from the basis (C1-1) to that of eq. (C2-1) is
in fact a group rotation which either rotates $\sqrt{2} P_1$ to $\widehat e_L$
 if
$M^2$ is positive or rotates $J$ to $\widehat e_L$ if $M^2$ is negative.  The
invariant metric as well as the Lie algebra structure is also
preserved by such a rotation.  Hence, \hfil\break
$[\widehat e_L, \widehat e_{T_1}] =
+(-) 2i\widehat e_{T_2}; [\widehat e_L, \widehat e_{T_2}] =
+(-) 2i\widehat e_{T_1}$ with the signs depending on the sign of
$M^2$.  Therefore
$[\widehat e_L A_\mu] = 2i (+(-) E_{T_{1\mu}}  \widehat e_{T_2}
+(-)  E_{T_{2\mu}}  \widehat e_{T_1}$.
 Rewriting eq. (C1-5)
$$
A_\mu = e_\mu P_+ + \lambda_\mu P_- + \beta_\mu J\leqno{(C1-5)}
$$
Gauss' law allows to express two gauge field components (say $\beta_\mu$
and $\lambda_\mu$) in terms of the third one $(e_\mu)$ and the
components of $\phi$. Indeed,
$$
\eqalign{
&\beta_\mu^{cl} = -{{i\partial_\mu \rho}\over\rho} + {\alpha \over \rho}
e_\mu\cr
&\lambda^{cl}_\mu = -{{i\partial_\mu\alpha}\over \rho} + {{M^2 +
\alpha^2}\over{2\rho^2}}e_\mu.\cr
}\leqno{(C2-3)}
$$
Thus
$$
\eqalign{
&E_{L_\mu} = {1 \over{\sqrt{M^2}}}({{M^2}\over \rho}e_\mu -
i\partial_\mu \alpha + i{{\alpha \partial_\mu\rho}\over\rho})\cr
&E_{T_{1\mu}} = {{\partial_\mu \alpha}\over{\sqrt{M^2 + \alpha^2}}}\cr
&E_{T_{2\mu}} = - {{\alpha \partial_\mu \alpha}\over{M\sqrt{M^2 +
\alpha^2}}} + {{\sqrt{M^2 + \alpha^2}}\over{\rho M}} \partial_\mu
\rho.\cr
}\leqno{(C2-4)}
$$
First note that the transverse fields are independent of $e_\mu$,
we will show that they transform as regular vectors under $H$, so they
can play the role of a zweibein. We will also show that the longitudinal
component $E_{L_\mu}$ transforms as the appropriate gauge field of the
residual gauge group $H$.

A gauge transformation in $H$ is defined by a gauge parameter
$$
\lambda(x) = \varepsilon(x)\widehat e_L\leqno{(C2-5)}
$$
and, under it, $A_\mu$ transforms as
$$
\delta A_\mu = \partial_\mu \lambda + [A_\mu, \lambda] = \widehat
e_L\partial_\mu \varepsilon + \varepsilon \partial_\mu \widehat e_L +
\varepsilon[A_\mu, \widehat e_L].\leqno{(C2-6)}
$$
The first term is by definition longitudinal while the last two are
transverse because
$$
\cases{Tr(\widehat e_L\partial_\mu\widehat e_L) = {1 \over
2}Tr\partial_\mu \widehat e^2_L = 0\cr
Tr\{\widehat e_L[A_\mu, \widehat e_L]\} = Tr(\widehat e_L A_\mu \widehat e_L) -
Tr(\widehat e_L \widehat e_L A_\mu) = 0.\cr}\leqno{(C2-7)}
$$
Recalling that $\widehat e_L = \phi/\sqrt{M^2}$ we realize that
Gauss' law implies the cancellation of the last two terms in eq. (C2-6)
and thus that $A_{\mu T}$ does not transform under $H$. This means that
$\vec E_{T_\mu} \cdot \vec{\widehat e}_T$ is invariant. Therefore $\vec
E_T$ transforms as $\vec{\widehat e}$ and thus as a regular vector under
$H$.

On the other hand
$$
\delta E_{L_\mu} = \partial_\mu \varepsilon\leqno{(C2-8)}
$$
showing that $E_L$ transforms as an $H$ gauge field.

Once proven that $E_{T_i}^a$ are 2-d space time vectors transforming as
vectors under $H$ - as zweibeins should do - we may define the metric
$g_{\mu\nu}$, invariant under the residual gauge group $H$, as

$$
g_{\mu\nu} = Tr ( ( E_{T_{1\mu}}  \widehat e_{T_1} +
E_{T_{2\mu}}  \widehat e_{T_2}) (E_{T_{1\nu}}  \widehat e_{T_1} +
E_{T_{2\nu}}  \widehat e_{T_2}) )
\leqno{(C2-9)}$$

Notice that the norms squared of the three basis elements defined in
eq. (C2-1) depend on M and on $\alpha$.  The signature: $[\widehat e_L^2,
\widehat e_{T_1}^2, \widehat e_{T_1}^2]$ is:  $[+1, -1, -1]$ if $M^2 >0$;
$[-1, -1, +1]$ if $M^2<0$ and $M^2 + \alpha^2 >0;$
$[-1, +1, -1]$ if $M^2<0$ and $M^2 + \alpha^2 <0$,
accordingly the definition of the metric in eq. (C2-9) is $g_{\mu\nu} =
E_{T_{1_\mu}}E_{T_{1_\nu}} +
E_{T_{2_\mu}}E_{T_{2_\nu}}$ if $M^2 >0$
 $g_{\mu\nu} =
E_{T_{1_\mu}}E_{T_{1_\nu}} -
E_{T_{2_\mu}}E_{T_{2_\nu}}$ if $M^2 <0$

On the other hand, from Gauss' law
$$
{{\partial_\mu\phi}\over M} = -[A_\mu, {\phi \over M}] = - [A_\mu, \widehat
e_L]\leqno{(C2-10)}
$$
meaning that in gauge space $\partial \phi$ is transverse, has the same
modulus as $\vec A_T$ and is perpendicular to it.

Therefore
$$
g_{\mu\nu} =
{{Tr(\partial_\mu\phi\partial_\nu\phi)}\over{2M^2}}.\leqno{(C2-11)}
$$
Let us comment on this result. We notice first that the induced metric is not
only locally invariant under $H$ but also globally under $G$.  Since the
surface $Tr \phi^2=M^2$ embedded in the 3 dimensional $\phi$ space
with invariant metric, has a constant curvature proportional to
$1 \over M^2$, the metric in eq. (C2-11) induced on the world sheet has a
constant, M independent curvature.  Notice that for $M^2 >0$ it
has Euclidean signature while Minkowski signature emerges for $M^2 <0$.

We could have defined the zweibeins, or the metric, by multiplying the
$E_T$ in eq. (C2-4) or $\partial_\mu \phi$ by an arbitrary function of M (the
only $H$ invariant function) to obtain any (possibly M dependent)
 constant value for the
curvature.

Let us now pause to consider what actually is the extent of the
classical promise for a formation of a metric.

As $M^2 \equiv {1 \over 2} tr \phi^2$ is constant, one may use it to classify
$\phi^a$ configurations. For a given value of $M^2$, all non constant
configurations lead, by the equations of motion, to induced world sheet
metrics which differ only by a gauge transformation, for non singular
metrics they differ only by a diffeomorphism as well. These metrics,
when invertible, lead all to the same constant curvature, that induced
by the group. Singular metrics lead to locally ill-defined curvatures,
for constant configurations, the metric vanishes globally as well.

Thus classically, up to diffeomorphism and to topological
considerations, there exists only one induced metric, the theory allows
the existence of a fixed metric, a metric with some singularities as
well as of no metric.

The quantum theory can either wash out or maintain that structure, it
may also add quantum fluctuations to the metric based configurations. We
will find out that the quantum theory does maintain the structure; however
as the theory is topological many of the equations of motion are valid
quantum mechanically limiting severly the allowed quantum fluctuations of
the induced metric. Put differently, pure 2-d quantum gravity is in any
case at best a topological theory, one can't expect much more than the
appearance of a space with very rigid curvature.

\vs
\n
{\bf C3.} {\it Quantum mechanical analysis-flat potential.}
\v
As we have stated several times, an intriguing structure that emerges in
topological theories is a flat potential even at the quantum level. One
may choose any wave functional obeying Gauss' law and inquire as to the
type of universe encoded in it.

We first consider physical states confined to a fixed value of $M^2$.
They can be classified in many respects according to the sign of $M^2$

Among all orbits a special role is played by the one defined by
$$
\phi^a = (0, 0, 0) ; M^2 = 0.\leqno{(C3-1)}
$$
This orbit consists of a single  $\phi$ configuration. It maintains the
full $SO(2, 1)$ gauge symmetry and all the world sheet is mapped into a
single point in target  space - the tip of the light-cone. By eqs. (C2-14)
(C2-17) this constant map does not induce a metric, the
theory is purely topological. The wave functional carries also no phase.
(In the $\Psi(A_i)$ picture, the choice of a singlet representation
allows all holonomies with equal amplitude). This wave functional carries some
analogy to the $\phi^a = 0$ special
point of flat potentials. It represents a group singlet and its choice
breaks no symmetry.

Symmetry is broken on the other orbits. Let's
first consider an orbit with $M^2 > 0$.

In a sense all these $M^2$ orbits are equivalent (just as different
values of $(\phi^a)^2$ in the case of a flat potential), they describe
two sheeted hyperboloids and the euclidean metric on these hyperbolids
is pulled back to the world sheet.

The support of the wave functional, as discussed in detail in section B, lies
in ${{SO(2, 1)}\over{T}}$, for any non-constant value of
$\phi^a$, eq. (C1-20) leads in target space to a metric which has a
Euclidean target space signature and to the same target space constant
curvature.

Fixed $\phi^a$ configurations on the other hand map the world sheet into
a single point target space, not allowing therefore any target space as
well as world sheet metric interpretation. Constant maps do not separate
between different space points. Their totality maps the full target
space, but one can't associated metric to any of them. For constant map
the phase of the wave functional associated with a target space area
also vanishes. Points for which a configuration $\phi(x)$ has vanishing
first derivative cannot participate in a metric space.

The wave functional describes a very rigid gravity. All non-constant
configurations are general coordinate equivalent (or equivalently gauge
equivalent) and thus describe closed loops on a single manifold of fixed
curvature (up to topological obstructions). The singular configurations
are gauge but not diffeomorphic equivalent and describe topological
pockets, their measure is not clearly defined and thus we can only state
their presence. The above is just the restatement of the relation
between gauge and co-ordinate transformations.

For $M^2 < 0$ the manifold described is the one sheeted hyperboloid with
a Minkowski metric induced by the group metric. For $M^2 = 0$ one obtains
the light like metric. A wave function which is a superposition of
different values of $M^2$ can also be considered, including the
superposition of different metric signatures (following from different
sign of $M^2$).

A contracted form of the group $SO(2, 1) \times U(1)$ was also given a
gravitational meaning [7], [8] (actually the euclidean group $E(2)$ was
extended). In fact it was suggested to relate it to a theory of the type
$$
\int d^2 x \sqrt g(\eta R - \Lambda)\leqno{(C3-2)}
$$
which has in particular zero curvature. Indeed as seen from eq. (B3-9)
the metric induced from the group is flat and thus the curvature
vanishes as in eq. (C3-2).

In [7], [8] a metric of the type of eq. (C1-6) modified by a
multiplication by a factor $(\phi^0)^{-1}$ was also constructed. That
metric was identical to that of a two dimensional black hole [15].

In a sense any symmetric tensor can be used to induce a metric however
the only $U(1)$ invariant type metrics we can construct are of the form
of a product of some function of $M^2$ time eq. (C1-20) and thus are all
non-singular.

Nevertheless, let's go along with the black hole
interpretation of the states and see the implications. Denote by $M^2$,
defined in eq. (B3-8) the mass of the black hole
and $\phi_D^0$ some $U(1)$ electric field. First note that it may be not
so easy to identify in an exact quantum mechanically solution a black
hole. After all the exact theory may have resolved the black hole
singularity as well as its other malises.

We remark that the Hilbert space connects only states labelled by $M$ it
seems that there is only one state for each $M$. If one wishes to
consider the spacial horizon as containing only 2 points the result
could confirm the counting of black hole entropy. On the other hand the
standard 2-d thermodynamic arguments lead to an entropy proportional to
$M$. In the absence of matter it is difficult to decide to which result
to attribute the entropy. The result
holds for a balck hole with a fixed value $C$ otherwise the number of
black holes becomes infinity.

The black hole is stable, adding to it a source, one obtains a new
configuration as shown in general in section B4.

In particular the ''mass'' of the black-hole (like any other electric
field) gets shifted in between the sources, as does the electric field
$\phi^0_D$. The shift is proportional to the group charge which can be
interpreted as particle momneta [7] in this case, echoing perhaps a
shift in the horizon expected for a black hole in that case [7].

Had we studied a theory such as $SO(2, 1) \otimes [SO(2, 1) \times
U(1)]$ one could have wave functionals that could describe a 2-d
Minkowski space times a 2-d black-hole in overall four dimensions.

\vs
{\bf C4.} {\it Target space aspects.}
\v
As physics is determined by $\Psi(\phi(x))$, $\phi(x)$ may be assigned a
role of a target space coordinate. This was done in particular in [7].
Moreover in a string picture $\phi(x)$ describes its configuration
in target space. To appreciate the consequence of such an association it
is maybe useful to consider first the case of compact $G$ as for example
$G = SU(2)$.

A basis for such a theory are configurations in which the string, in its
entirety, must lie on a sphere with a fixed integer radius (constant
$M^2$).
This is not the conventional way a string moves on a
manifold. In string theory there are limitations on the allowed
configurations of a classical string but no regions in target space are
excluded. While here only regions which lie on a shell (whose radius is
positive) can be visited by the string.
The union of these platonic orbits does not span
the full $R^3$. While it is true we gain a target space of a metric
meaning, what is obtained is a very limited target space. This
limitation reflects the very scarce possibilities for fluctuation.

For the non compact case, $G = SO(2, 1)$, the space spanned consists of
the two sheeted hyperboloid, with an Euclidean metric and a quantization
condition $(M^2 > 0)$, a one sheeted hyperboloid $(M^2 < 0)$, a light
cone and its tip $(M^2 = 0)$.

One may decide to assign the $\phi$ field
the role of a conjugate momentum. In that case the target space role
shifts to the manifold of gauge configurations. It is not by changing
language that the phases will change.

In the $\phi$ picture for a fixed value of $M^2$, the wave
functional is a pure phase and of the form $\Psi(\phi^a) = exp i
S(\phi^a)$, thus assigning equal probability to any configuration
$\phi^a$. This reflects the equivalence of configurations related by a
gauge transformation.

The concept of an area does however reside in the phase as we have shown
in the former section. In the calculation of expectation values of operators
depending only on
$\phi^a$ the wave functional will contribute as the identity.

In a more general wave functional of the form
$$
\Psi(\phi^a) = exp iS(\phi^a)F(M^2)
$$
amplitudes for different values of $M^2$ will not interfere.

There is however one geometrical feature which will emerge.

Any given configuration $\phi^a$ corresponds to a target space metric,
which has already made its imprint by allowing the appearance in the
phase of the wave functional of the area bounded by the loop $\phi^a$ on
the surface defined by $M^2$. This same metric can be used to define a
corresponding length of the circumference of the world sheet circle. The
value of this length does vary with $\phi^a$, that is by definition is not
locally
$SO(2, 1)$ invariant.

Nevertheless it will enable us to set a handle on the emergence of the
rigid but existing metric, by evaluating an order parameter.
\vs
\n
{\bf C5.} {\it An order parameter.}
\v
In the Hamiltonian formulations, what is left of the metric is just the
$g_{11}$ component, which is locally $H$ but only globally $G$ invariant.

Nevertheless following a general suggestion [6] for an order parameter
for gravity we can mention the induced metric in full.

Let's first review the suggestion.

Assume one is given a quantum gravity described by some action $S$ and
one is given a loop (defined by $x^\mu (s)$ $ 0 \leq s \leq 1$) in space. We
are
essentially instructed to calculate the length of the loop averaging
over the metrics.

One defines $Z(\mu)$ by,

$$
Z(\mu) = <exp
-\mu \int_0^1 ds\sqrt{g_{\mu\nu}{dx^\mu \over ds} {dx^\nu \over ds}}> =
\int D\chi \int Dg_{\mu\nu} exp(-S) exp(-\mu \int_0^1
ds\sqrt{g_{\mu\nu}{dx^\mu \over ds} {dx^\nu \over ds}})\leqno{(C5-1)}
$$
$\chi$ representing possible other matter fields.  Since eq. (C5-1)
contains integration over all metrics, $Z(\mu)$ is independent  on the labeling
of the position of the loop.  We can redefine $Z$, up to an infinite $\mu$
independent,
constant, to include an integration over all these
positions.

$$Z(\mu) = \int D\chi \int Dg_{\mu\nu} exp(-S)\int Dx^\mu(s) exp(-\mu
\int \sqrt{g_{\mu\nu}dx^\mu dx^\nu}).\leqno{(C5-2)}$$
Using the standard replacement (which is rather redefinition of $Z(\mu)$ than
an identity) one obtains,

$$Z(\mu) = \int D\chi Dg_{\mu\nu} exp(-S) \int Dx^\mu(s) De(s)
 exp(-\mu \int_0^1  ({{g_{\mu\nu}}(x^\mu(s))\over e(s)}
{{d x^\mu}\over{ds}}{{d x^\nu}\over{ds}} + e(s))ds) \leqno{(C5-3)}$$

Gauge fixing $e$ to a constant value $L$ given for each $e(s)$ by:

$$
\int_0^1 e(s)ds = L\leqno{(C5-4)}
$$
and assuming no ghosts (they actually renormalize the one dimensional
cosmological constant), one has
$$
Z(\mu) = \int  dL \int D\chi Dg exp(-S) \int Dx^\mu(s)
 exp(-\mu L- {1 \over L}
 \int_0^1
ds{g_{\mu\nu}{dx^\mu \over ds} {dx^\nu \over ds}})\leqno{(C5-5)}
$$
so that the Laplace transform $Z(L)$ is given by:
$$
Z(L) =< \int D x^\mu(s) exp - {1 \over L} \int
ds{g_{\mu\nu}{dx^\mu \over ds} {dx^\nu \over ds}}>\leqno{(C5-6)}
$$
 which is the average of the propagator of a quantum particle moving
on the fluctuating manifold.

It was suggested [6] to search for the small L dependence of eq. (C5-6) in a
general model and parametrize it by $L^{-D \over 2}$ at small L.  D
would then represent the effective dimensionality of the quantum geometry.
$D=0$ would indicate a topological phase.

In our case we can pick the test loop of eq. (C5-1) to be a fixed time slice of
the manifold and calculate the expectation value of eq. (C5-1) in the
Hamiltonian
formalism using the explicit wave functionals.  For a wave functional
corresponding to a fixed value of M, eq. (C5-1) becomes,

$$Z(\mu) = \int D\phi(x)|\Psi(\phi (x))|^2  exp(-\mu
\int\sqrt{g_{11}(x)}dx) \leqno{(C5-7)}$$

$\Psi$ being a pure phase drops out of the expression, substituting
for $g_{11}$ the induced metric eq. (C2-11)\footnote{(*)}{by
Gauss' law eq. (B2-6) the replacement $\partial_x\phi
\leftrightarrow - [A, \phi]$ is valid also in quantum mechanics, forming
the composite operator assumes no short distances subtellies otherwise
it is
a definition of $g_{11}$} one obtains

$$
Z(\mu) = \int D\phi(x) exp(-\mu \int
\sqrt{B_{ab}(\phi^\mu)d\phi^a d\phi^b}).\leqno{(C5-8)}$$
where $B_{ab} d\phi^a d\phi^b = {1\over{M^2}}
Tr (d\phi d\phi)$.
Using again the standard replacement

$$
Z(\mu) = \int D\phi(x) De(x)
 exp(-\mu \int ({{B_{ab}}(\phi^\mu(x))\over e(x)}
{{d \phi^a}\over{dx}}{{d \phi^b}\over{dx}} + e(x))dx) \leqno{(C5-9)}$$
and the above gauge fixing leads to the Laplace transform,

$$
Z(L) =\int D \phi(x) exp (- {1 \over L} \int_0^1
dx B_{ab}(\phi(x)){{d\phi^a} \over {dx}} {{d\phi^b} \over {dx}})
\leqno{(C5-10)}
$$
which describes the motion of a quantum particle on the target space,
the two-dimensional orbit endowed with the metric $B_{ab}$.
The short distance behavior of this propagator corresponds to $D=2$,
the dimension of target space.  For the wave functional representing
the singlet orbit one obtains $D=0$.

For the one and the two sheeted hyperboloids endowed with the group metric the
exact result for $Z(L)$ may be obtained [16]; the highest order in
${1 \over L}$ reveals the dimensionality of the manifold while non
leading orders will provide other properties as the curvature. Thus
gauge invariant information (such as $D$ and $R$) can be extracted from
an object which is not itself fully gauge-invariant).

In the gauge field representation we have also evaluated the wave functional in
the
presence of external charges noticing that the topological theory forms
tensionless strings\footnote{(**)}{The inverse property, i.e. that tensionless
strings
provide topological theories, has recently been advocated [18].}. We have
also noted that these strings will gain tension once the theory is
deformed by adding a perturbation $tr\phi^2$. From the gravitational
point of view the modified theory has less symmetry than the topological
theory, it is a function of the world sheet area. It does have
however more symmetries than a usual GCI theory as no metric needs to be
introduced in order to obtain invariance under area preserving
transformations.

It is interesting that this intermediate realization of GCI has an
identical Hilbert space to that of the topological realization, where as
we have discussed the $tr\phi^2$ term acts as a ''magnetic'' field in
removing the energy degeneracy of the states.

One could also deform the $\phi F$ model and remain topological. As
shown in [17] the addition of an action such as, $\mu \int d^2 x\sqrt g
R(tr \phi^2)$, spans one direction in the moduli space of topological
theories in which $\phi \cdot F$ theories are members. One call add
also other group invariant functions of $\phi$ to parametrize more
directions in moduli space. For such theories the value of the one point
functions such as $tr \phi^2$, would be deformed (from the gauge theory
point of view) but the energy of all states would remain zero.
In [17] it was shown that some topological theories with a finite number
of observables can be rewritten as string theories of other topological
thoeries. The same is true for the $\phi*F$ theory, and stems
from the it's relation to the Chern-Simmons theory.
It was shown [19] that a $CS$ theory with an $SU(N)$ group can be rewritten as
a string theory of open string on various Calabi-Yau manifolds . The level
$k$ of the $CS$ theory serves as a dilaton expansion parameter.
In the limit of very large $k$ one obtains on the one hand the $2$-d
$\phi*F$ theory as an effective string
theory and on another hand the string theory itself is the large $k$ limit
of the Calabi-Yau theories. This is an example of an effective field theory of
a theory with an infinite though discrete number of observables.

\vs
\n
{\bf C6.} {\it Product group and a possibility for scale generation.}
\v
In the commutation relation of eq. (C1-1) a scale $\Lambda$ appeared
explicitly and has been set to one. One could consider in general $\phi
F$ theories for very large and maybe even infinite groups, one could
further obtain effective theories with reduced symmetry, by considering
special orbits along the flat potential. However as the group rank is
conserved (the scalar $\phi$ is in the adjoint representation) the
reduced symmetry has to contain a bunch of $U(1)$ like factors. One
could also consider product of groups, for the gravitational
interpretation this would allow having products of various universes.

There is one issue which actually leads us to product groups and that is
the issue of scale invariance, one would expect topological theories to
be scale invariant, however for Einstein like realization of GCL one
could have in principle both scale invariant realizations (for example
if conformal gravity made sense) and non scale invariant realizations
(manifest by Newton's constant).

For the gravitational models of the $\phi \cdot F$ type a scale seems to
be explicitly involved in the commutation relations (eq. C1-1).
Following of the idea [20] that the Poincar\'e and De-Sitter groups
may offer as spontaneously broken realization of a conformal theory we
gauge the group $SO(2, 2)$ which is the anomaly free sub-group of the
infinite dimensional conformal group in 2 dimensions. We may eventually
wish also to construct a $\phi \cdot F$ theory for the full Virasoro
group.

For $SO(2, 2)$ one can construct all commutation relation such that no
scale appears, moreover as $SO(2, 2)$ is actually locally a product of
two $SO(2, 1)$ factors, we use the following relations,
$$
[J + D, P_+] = -2iP_+ \qquad [J - D, K_+] = - 2iK_+
$$
$$
[J + D, K_-] = 2iK_- \qquad [J - D, P_-] = 2iP_-
$$
$$
[P_+, K_-] = 2i(J + D)\qquad [K_+, P_-] = 2i(J - D)
$$
$P_+, P_-$ are the translation operators, $K_+, K_-$ the operators
related to special conformal transformations, $J$, is the Lorentz boost
and $D$ is the dilatation operator.

No scale is involved in these commutations relations. The two $\phi^i$,
however, have components of different dimensionality so
that the adimensional Casimirs
$$
M^2_i = 2\sigma_i \rho_i - \alpha^2_i \qquad (i = 1, 2)
$$
are made up by opposite dimensional $\rho_i$ and $\sigma_i$ quantities.
A non zero field configuration would therefore provide a scale. The wave
functionals of this model, treated as a $G \times G$ one, may describe
universes varying from a point to a product of two Euclidean
two-sheeted hyperbolids.

In the point solution, for which the wave functional is a product of
the wave functionals representing the singlet in both $SO(2, 1)$'s,
clearly no scale is generated. The same is true for constant $(\phi_1,
\phi_2)$ configurations as the group invariant quantities that may be
constructed with them (such as $tr\phi^2_1 + tr\phi^2_2)$ introduce no
scale.

However as the wave functionals not representing the group singlet
necessarily have support on non constant configurations, the
dimensionfull fields $\rho_i(x)$ and $\sigma_i(x)$ do provide a scale.
\v
\n
{\bf Concluding Remarks.}

Induced metric theories have a general coordinate invariance without
introducing ab initio a metric - and thus a graviton - and as such  they
fall in the cathegory of classical topological systems.

Field quantum fluctuations generate a metric that will at its time
fluctuate at higher orders of this semiclassical approach. An approach
which is however badly suited to reveal [21] how the underlying topology
inhibits large metric fluctuations.

We have exploited the exact solvability of the $2$-d $\phi F$ theory to
find that the induced metric generated in the semiclassical approach
emerges also at the quantum level. There exist a quantum state for which
the gauge symmetry is unbroken and which corresponds to a non metric
universe. For other states the gauge symmetry is broken and a metric
appears. Some quantization conditions of the
invariant Casimirs limit the possible classical field configurations.
The induced gravity shows very rigid properties
(constant curvature) that may be assigned to the lack of metric
fluctuations in $2$-dtheories.

We recognize that the systems we study - while simple enough to exibit at a
full quantum level the generation of a metric - perhaps too
simple to shed light on how induced metric theories succeed in taming
the wild fluctuations of quantum gravity. Indeed, the models studied and
solved exactly have too few degrees of freedom to allow for metric
fluctuations.

\v
\centerline{\bf Acknowledgement}

We wish to thank E.Witten, T.Banks and Ch.Grosche for discussions;
E. Rabinovici has been partially supported by the Israel Science Foundation,
the
Bi-National American-Israeli Science Foundation, the
NSF Grant \#PHY92-45317 and The Ambros Monell Foundation; S.Elitzur by
the Israel Science Foundation; D.Amati by the EEC Grant SC1*-CT92-0792.
The efficient secretarial help of Claudia Parma and Erika Zynda are gratefully
acknowledged.
\vs
\centerline{\bf References}
\vs
\item{[1]} A.D.Sakharov, Sov. Phys. Dokl. 12 (1968), 1040; D.Amati and
G.Veneziano, Nucl. Phys. B204 (1982), 451.
\item{[2]} E.Witten, Comm. Math. Phys. 117 (1988), 353; Nucl. Phys.
B311 (1988/89) 46.
\item{[3]} E.Witten, Nucl. Phys. B311 (1988/89) 46.
\item{[4]} A.Chamseddine and D.Wyler, Phys. Lett. B228 (1989) 75;
Nuclear Physics B 340 (1990) 595; K.Isler and C.A.Trugenberger, Phys.
Rev. Lett. 63 (1989) 834; D.Montano and J.Sonnenschein Nucl. Phys.
B324(1989) 348.
\item{[5]} E.Witten, Commun. Math. Phys. 141 (1991) 153.
\item{[6]} A.M.Polyakov, Lecture presented at Les-Houches 1992, Princeton
University, preprint.
\item{[7]} H.Verlinde, String theory and quantum gravity, ICTP, 1991;
World Scient. p. 178 1992, Sixth Marcel Grossman Conference Meeting on
General Relativity M.Sato, ed. (World Scientific, Singapore, 1992).
\item{[8]} D.Cangemi and R.J.Jackiw, Phys. Rev. Lett. 69 (1992)
233; R.Jackiw, MIT preprint, CTP 2105 (1993).
\item{[9]} E.Witten, Commun. Math. Phys. 121 (1989) 351.
\item{[10]} S.Elitzur, G.Moore, A.Schwimmer and N.Seiberg, Nucl.
Phys. B326 (1989) 104.
\item{[11]} A.Alekseev, L.Fadeev and S.Shatashvili, Journal of Geometry
and Phys. 5 (1989) 391.
\item{[12]} G. 't Hooft, Nucl. Phys. B79 (1974) 276; A.M.Polyakov,
JEPT Lett. 20 (1974) 194.
\item{[13]} N. Seiberg, Notes on Quantum Liouville Theory and Quantum
Gravity. In: Common Trends in Mathematics and Quantum Field Theory, Proc. of
the
1990 Yu` Progress of Theo. Phys. Sup.  102
319 (1990).Edited by T.Eguchi, T. Inami and T. Miwa.   To appear in the Proc.
of the
Cargese meeting, Random Surfaces, Quantum Gravity and Strings, 1990.
RU-90-29.
\item{[14]} C.Teitelbaum, Phys. Lett. B126 (1983): in Quantum theory
of gravity, S.Christensen ed. (Adam Hilger, Bristol, 1984); R.Jackiw,
ibid; Nucl. Phys. B252 (1985) 343.
\item{[15]} K. Bardakci, M.Crescimanno and E.Rabinovici, Nucl. Phys.
B349; S.Elitzur, A.Forge and E.Rabinovici, Nucl. Phys. B359 (1991)
581; E.Witten, Phys. Rev. D44 (1991) 314.
\item{[16]} C.Grosche and F.Steiner, Ann. Phys. 182 (1988) 120;
C.Grosche, in preparation.
\item{[17]} S.Elitzur, A.Forge and E.Rabinovici, Nucl. Phys. B359 (1991)
581.
\item{[18]} J.Isberg, V.Lindstrom, B.Sundbourg and G.Theodoridis,
VSITP-93-12; A.Schild, Phys. Rew. D16 (1977) 1722.
\item{[19]} E.Witten, IAS Preprint 92/45, 1992.
\item{[20]} S.Fubini, Nuovo Cimento 34A, 521 (1976).
\item{[21]} cf. however D.Amati and J.Russo, Phys. rew. Lett. B248 (1990),
44-50.

\bye